\documentclass[reprint,
superscriptaddress,
 amsmath,amssymb,
 aps,
]{revtex4-2}

\usepackage{graphicx}
\usepackage{dcolumn}
\usepackage{bm}
\usepackage[colorlinks=true, linkcolor=blue, urlcolor=blue,citecolor=blue]{hyperref}

\usepackage{xcolor}
\begin{document}

\preprint{APS/123-QED}

\title{Tunable magnon band topology and magnon orbital Nernst effect in noncollinear antiferromagnets}

\author{D. Quang To}%
\email{quangto@udel.edu}
\affiliation{Department of Materials Science and Engineering, University of Delaware, Newark, Delaware 19716, USA}

\author{Dai Q. Ho}
\affiliation{Department of Materials Science and Engineering, University of Delaware, Newark, Delaware 19716, USA}
\affiliation{Faculty of Natural Sciences, Quy Nhon University, Quy Nhon 55113, Vietnam}

\author{Joshua M. O. Zide}%
\affiliation{Department of Materials Science and Engineering, University of Delaware, Newark, Delaware 19716, USA}%

\author{Lars Gundlach}
\affiliation{Department of Chemistry and Biochemistry, University of Delaware, Newark, Delaware 19716, USA}

\author{M. Benjamin Jungfleisch}
\affiliation{Department of Physics and Astronomy, University of Delaware, Newark, Delaware 19716, USA}

\author{Garnett W. Bryant}
\affiliation{Nanoscale Device Characterization Division, Joint Quantum Institute, National Institute of Standards and Technology, Gaithersburg, Maryland 20899-8423, United States}
\affiliation{University of Maryland, College Park, Maryland 20742, USA}

\author{Anderson Janotti}
\affiliation{Department of Materials Science and Engineering, University of Delaware, Newark, Delaware 19716, USA}

\author{Matthew F. Doty}%
 \email{doty@udel.edu}
\affiliation{Department of Materials Science and Engineering, University of Delaware, Newark, Delaware 19716, USA}

\date{\today}

\begin{abstract}{} 
We theoretically investigate the intrinsic magnon orbital Nernst effect (ONE) in noncollinear antiferromagnets with Kagom\'e spin systems. Our analysis reveals that an externally applied magnetic field induces topological phase transitions in the magnonic system, characterized by the closing and reopening of the band gap between distinct magnon bands. These transitions enable tunable control of the magnon orbital Nernst effect with applied magnetic field, with a pronounced enhancement in magnon orbital Nernst conductivity near the phase transition points. This tunability presents a promising direction for experimental detection of the magnon ONE.
\end{abstract}

\maketitle

\section{Introduction}

Magnons, which are collective excitations of spins in magnetic materials, have emerged as a focus of intense research \cite{Chumak2015, Pirro2021, Flebus2024} because they possess remarkable properties, including dissipationless propagation and nontrivial topology, that create opportunities for their use in low-energy information processing and high-speed device operation. \cite{Shindou2013, Mook2014a,Mook2014b, Mook2015, Laurell2018, Lu2019, Mook2019, Mcclarty2022,Zhuo2023,Yu2024} Moreover, magnons can hybridize with a variety of quasiparticles like photons \cite{Bhoi2019,Golovchanskiy2021}, electrons \cite{Carpene2008,Rohling2018,Kristian2021}, phonons \cite{Park2019, Mai2021,To2023b, Manley2024}, plasmons \cite{To2022,Costa2023, To2023a, Dyrdal2023}, and excitons \cite{Bae2022,Wang2023,Diederich2023}, which provides many opportunities for both scientific exploration and technological innovation. Historically, studies of magnon transport have centered on spin-based phenomena such as the spin Seebeck effect \cite{Xiao2010,Rezende2014,Rezende2016} and the magnon spin Nernst effect \cite{Cheng2016,Zyuzin2016,Li2020,Park2020, To2024,Cui2023}, which depend on the spin degree of freedom in magnons. Recent research, however, has revealed that the orbital moment of magnons offers additional opportunities in magnetic systems \cite{Matsumoto2011a, Matsumoto2011b, Neumann2020, Fishman2022, Fishman2023a, Fishman2023b, Zhang2020im,Alahmed2022,Huang2024, To2025, Go2024}.

\begin{figure}
\centering
\includegraphics[width=0.5\textwidth]{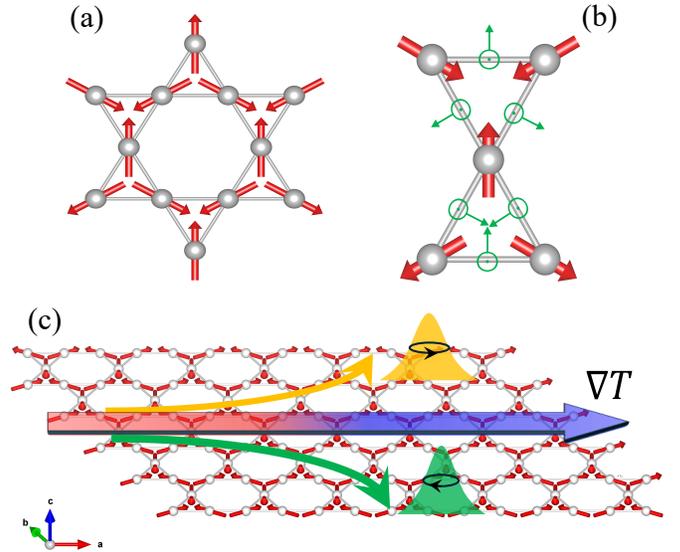}
\caption{(a) The Kagom\'e spin lattice of potassium iron jarosite and (b) Dzyaloshinskii-Moriya vectors (green) with both in-plane (arrows) and out-of-plane ($\odot$) components. (c) Schematic view of the magnon ONE in a Kagom\'e spin system where the transverse flow of magnons carrying opposite out-of-plane orbital moment is induced by a temperature gradient $\nabla T$ along the longitudinal direction.}
\label{Scheme}
\end{figure}

The magnon orbital Nernst effect (ONE) describes the generation of a transverse orbital angular moment current carried by magnons in response to a thermal gradient \cite{Zhang2019,To2025,Go2024}. Research into this effect can not only advance our understanding of orbital physics in magnetic systems, but also create technology opportunties. For example, the magnon ONE bridges the disciplines of spintronics, magnonics, and orbital transport, unlocking new possibilities for manipulating spin and orbital degrees of freedom in solid-state systems. Similarly, by introducing a mechanism to generate orbital angular moment without charge transport, the magnon ONE opens doors to innovative spin-orbitronic and magnon-based functionalities that could drive the development of energy-efficient spintronic devices, thermal management solutions, and next-generation quantum materials. 

\begin{figure}
\centering
\includegraphics[width=0.5\textwidth]{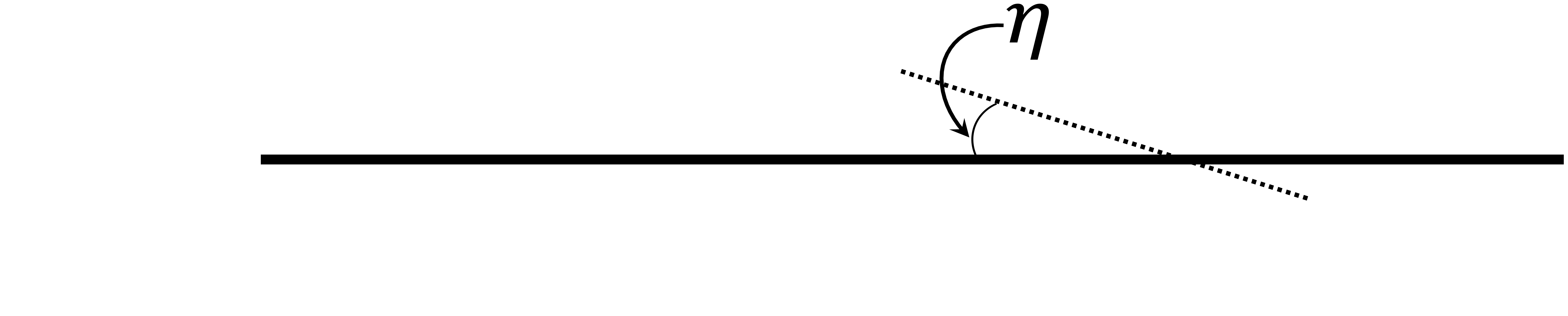}
\caption{Side view of a monolayer Kagom\'e spin lattice of potassium iron jarosite, illustrating weak ferromagnetism arising from a small spin canting angle, $\eta$.}
\label{FIG6}
\end{figure}

Previous investigations of the magnon ONE have focused on magnetic materials with honeycomb lattices \cite{Go2024, To2025, An2024}, demonstrating that this effect can occur even in systems without spin-orbit coupling (SOC). Similar to the orbital Hall effect of electrons in light metals, the magnon ONE does not inherently depend on SOC. However, experimentally detecting the magnon orbital moment through measurable quantities such as magnetization or electric polarization does require SOC or related interactions like magnon-phonon coupling. Moreover, previous studies suggest that an externally applied magnetic field has little impact on the magnon ONE, at least for the systems they considered \cite{To2025,Go2024}.

In this paper, we explore the magnon orbital Nernst effect in noncollinear antiferromagnets with Kagom\'e lattices. We show that noncollinear systems exhibit magnon band topology that can be tuned by externally applied magnetic fields, creating the opportunity to control magnon orbital moment transport properties. The paper is organized as follows: In Section \ref{Model}, we present the spin Hamiltonian model used to obtain the magnon dispersion in the Kagom\'e jarosite lattice through the spectral function. The Berry curvature and Chern number formulas are presented to describe the nontrivial topological magnon bands in the Kagom\'e jarosite lattice with finite Dzyaloshinskii-Moriya interaction (DMI). This section concludes with an introduction of the magnon orbital Nernst effect, employing linear response theory. In Section \ref{MONE}, we provide a detailed discussion of the magnon orbital Nernst effect in KFe$_{3}$(OH)$_{6}$(SO$_{4}$)$_{2}$, demonstrating its tunability as a function of the externally applied magnetic field. The connection between the tunability of the magnon orbital Nernst effect and the topological phase transition of magnon bands under the applied magnetic field is thoroughly examined. Sect.~\ref{Conclusion} summarizes the key findings.

\begin{figure*}
\centering
\includegraphics[width=0.8\textwidth]{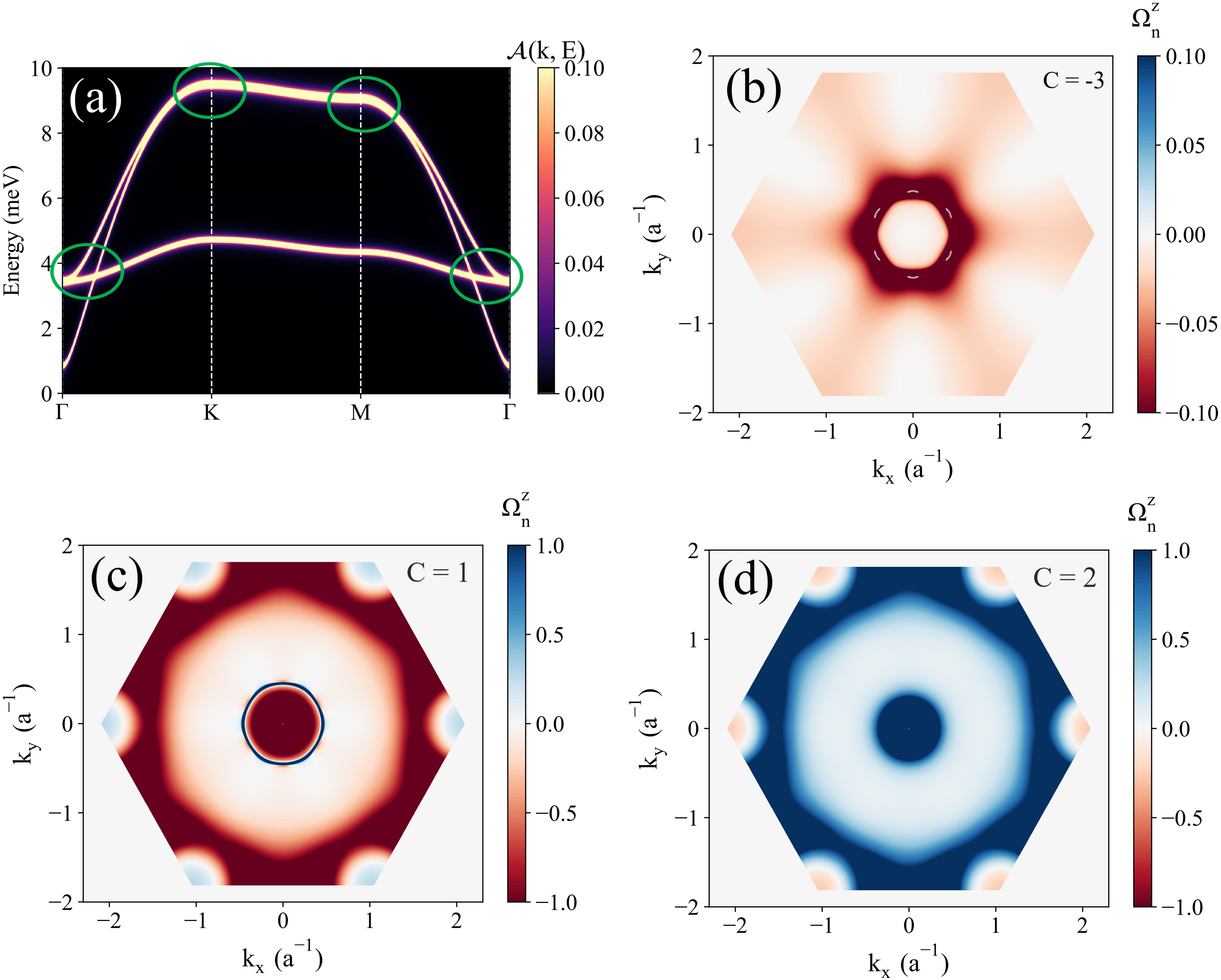}
\caption{(a) The spectral function $\mathcal{A}\left(k,E \right)$ calculated with magnon lifetime $\tau \approx 10^{-10}s$ \cite{Li2020b} showing the magnon dispersion in noncollinear AFM along the symmetric path $\Gamma-K-M-\Gamma$ in the absence of externally applied magnetic field ($B_{z}=0$). The green circles highlight the regions where anticrossing points emerge between two distinct bands. The Berry curvature of the lowest (b), middle (c), and upper (d) bands within the Brillouin zone along with the Chern number for each band. (b-d) are also calculated for $B_{z} = 0$.}
\label{FIG1}
\end{figure*}

\begin{figure*}
\centering
\includegraphics[width=1\textwidth]{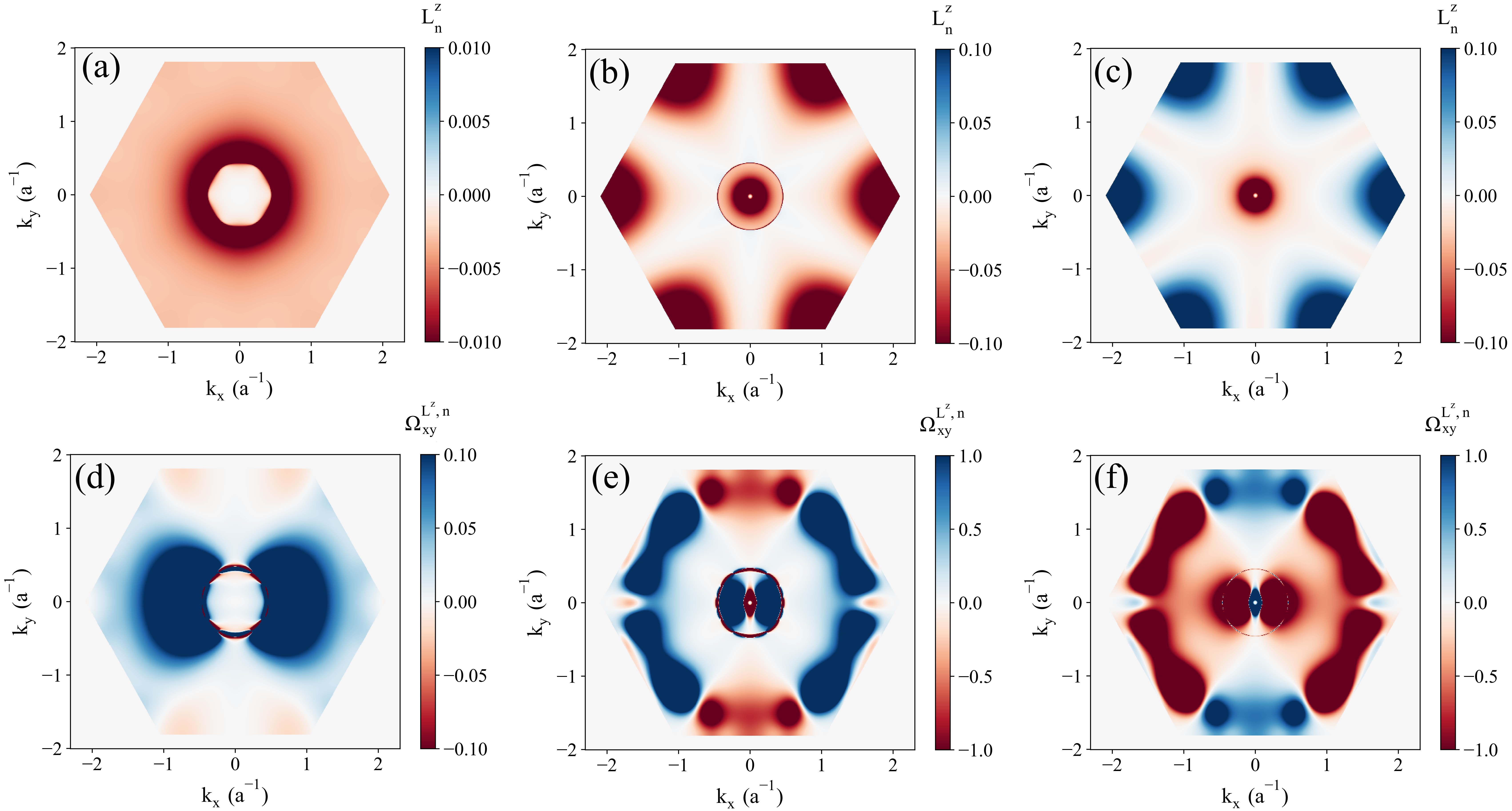}
\caption{The intra-band OAM (a-c) and orbital Berry curvature (d-f) of magnon bands in the Brillouin zone of KFe$_{3}$(OH)$_{6}$(SO$_{4}$)$_{2}$. (a,d) show the lowest energy band, (b,e) show the middle band, and (c,f) show the upper band. The calculations are performed for $B_{z} = 0$.}
\label{FIG2}
\end{figure*}

\section{Model and method} \label{Model}

\subsection{Model Hamiltonian for noncollinear spin systems}

In this work, we investigate the jarosite compound, taking as our example KFe$_{3}$(OH)$_{6}$(SO$_{4}$)$_{2}$, a material that crystallizes in a layered structure. The Fe ions in this compound form a two-dimensional Kagom\'e lattice characterized by a network of corner-sharing triangles. This unique arrangement is known to induce geometrical frustration in magnetic systems~\cite{Grohol2003, Grohol2005, Yildirim2006, Matan2006}. In geometrically frustrated spin systems, the energies associated with different spin interactions often compete, making it impossible to simultaneously minimize all interactions. This frustration arises from the inherent geometry of the lattice and can lead to unconventional spin structures and dynamics at low temperatures. For instance, the absence of conventional long-range magnetic order and the presence of potential quantum fluctuations make jarosite a promising candidate for studying quantum spin-liquid states \cite{Matan2006}. 

Figure \ref{Scheme}(a) illustrates a possible N\'eel state for an antiferromagnetic Kagom\'e lattice. In the classical limit, all ground states adopt a structure in which the angle between each nearest neighbor spin pair is about $120^{0}$. Specifically, Fig.~1(a) shows a $\boldsymbol{q}=0$ type ordering of Kagom\'e jarosite with positive chirality \cite{Grohol2005, Yildirim2006, Mondal2021}. At low temperature, typically below the N\'eel temperature ($T_{N}$, which is $65~K$ for potassium iron jarosite) the magnetic interactions in this lattice can be well described by the following Hamiltonian \cite{Matan2006, Laurell2018,Lu2019,Li2020}:

\begin{widetext}
\begin{align}
    H = J_{1}\sum_{\left\langle i, j \right\rangle} \boldsymbol{S}_{i} \cdot \boldsymbol{S}_{j}  + J_{2}\sum_{\left\langle\left\langle i, j \right\rangle\right\rangle} \boldsymbol{S}_{i} \cdot \boldsymbol{S}_{j}  + \sum_{\left\langle i, j \right\rangle} \boldsymbol{D}_{ij} \cdot \left( \boldsymbol{S}_{i} \times \boldsymbol{S}_{j} \right) - g\mu_{B}B_{z}\sum_{i}S_{i}^{z}
    \label{Hamilton}
\end{align}
\end{widetext}
where the first and second terms represent the nearest-neighbor and next-nearest-neighbor antiferromagnetic interactions, respectively. The third term describes the Dzyaloshinskii-Moriya interaction, characterized by the DM vector $\boldsymbol{D}_{ij}$ that includes both in-plane ($D_{p}$) and out-of-plane ($D_{z}$) components for the bond $(i,j)$. Here the out-of-plane $D_{z}$ stabilizes the $120^{0}$ coplanar $\boldsymbol{q} = 0$ spin structure. When $D_{z} = 0$, the coplanar structure can still be stabilized by the next-nearest neighbor interaction $J_{2}$. In contrast, the in-plane $D_{p}$ breaks both the mirror symmetry with respect to the kagome plane and the global spin rotation symmetry. This in-plane component induces a canting of the spins out of the plane, resulting in a weak ferromagnetic phase along the c-direction (see Fig.~\ref{FIG6}). The canting angle $\eta$ is determined by minizing the classical energy of the spin system in a Kagom\'e AFM, which is given by
 \begin{align*}
     \frac{E\left( \eta \right)}{NS^{2}} =& \frac{1}{2}\left(J_{1}+J_{2} \right) \left[3cos\left( 2\eta\right) - 1  \right]\\
     & - \sqrt{3}D_{z}cos^{2}\left(\eta \right)  - \sqrt{3}D_{p}sin\left(2\eta \right) - \frac{sin\left( \eta \right)g\mu_{B}}{S}B_{z}   
 \end{align*}
This energy is minimized ($\frac{1}{NS^{2}} \frac{\partial E\left(\eta \right)}{\partial \eta}=0$) when 
\begin{align}
    \eta = \frac{1}{2}tan^{-1}\left[\frac{-2D_{p}}{\sqrt{3}\left( J_{1} + J_{2}\right)-D_{z}} \right]
\end{align}
in the absence of applied magnetic field. When an applied magnetic field is present, this angle is given (for small angle $\eta$ and to first order) by:
\begin{equation}
    \eta  = \frac{-g\mu_{B}B/S - 2\sqrt{3}D_{p}}{6\left(J_{1} + J_{2}\right)- 2\sqrt{3}D_{z}} 
    \label{cantingangle}
\end{equation}
By applying the Holstein-Primakoff transformation \cite{Holstein1940,Vogl2020} to the spin operators, truncated to linear order, and performing a Fourier transformation, we obtain the Hamiltonian in Eq.\eqref{Hamilton} expressed in its second-quantized form, which incorporates the canting angle $\eta$, as defined in Eq.\eqref{cantingangle}, and is formulated in the basis $\Psi_{\boldsymbol{k}} = \left(b_{1,\boldsymbol{k}}, b_{2,\boldsymbol{k}}, b_{3,\boldsymbol{k}}, b_{1,-\boldsymbol{k}}^{\dagger}, b_{2,-\boldsymbol{k}}^{\dagger}, b_{3,-\boldsymbol{k}}^{\dagger}\right)$. The Hamiltonian takes the form: $H = \frac{S}{2}\sum_{\boldsymbol{k}}\Psi_{\boldsymbol{k}}^{\dagger}H_{\boldsymbol{k}}\Psi_{\boldsymbol{k}}$ where
\begin{align}
    H_{\boldsymbol{k}} = \begin{pmatrix} A_{\boldsymbol{k}} &B_{\boldsymbol{k}}\\
    B_{\boldsymbol{k}} &A_{\boldsymbol{k}}^{*}
    \end{pmatrix}
    \label{Hsecond}
\end{align}
Here 
\begin{widetext}
  \begin{align}
    A_{\boldsymbol{k}} = \begin{pmatrix} 2\left[\Delta_{1}^{(0)} + \Delta_{2}^{(0)} - \frac{g\mu_{B}sin\left(\eta \right)}{2S}B_{z}\right] & \Delta_{1}cos\left(k_{3}\right) + \Delta_{2}cos\left(p_{3} \right)  & \Delta_{1}^{*}cos\left(k_{2}\right) + \Delta_{2}^{*}cos\left(p_{2} \right) \\
     \Delta_{1}^{*}cos\left(k_{3}\right) + \Delta_{2}^{*}cos\left(p_{3} \right) &2\left[\Delta_{1}^{(0)} + \Delta_{2}^{(0)} - \frac{g\mu_{B}sin\left(\eta \right)}{2S}B_{z}\right]  & \Delta_{1}cos\left(k_{3}\right) + \Delta_{2}cos\left(p_{3} \right) \\
     \Delta_{1}cos\left(k_{2}\right) + \Delta_{2}cos\left(p_{2} \right) &\Delta_{1}^{*}cos\left(k_{1}\right) + \Delta_{2}^{*}cos\left(p_{1} \right) &2\left[\Delta_{1}^{(0)} + \Delta_{2}^{(0)} - \frac{g\mu_{B}sin\left(\eta \right)}{2S}B_{z}\right]
    \end{pmatrix}
\end{align}  
\end{widetext}
and
\begin{widetext}
  \begin{equation}
    B_{\boldsymbol{k}} = \begin{pmatrix} 0 & \Delta_{1}^{\prime}cos\left(k_{3}\right) + \Delta_{2}^{\prime}cos\left(p_{3} \right)  & \Delta_{1}^{\prime}cos\left(k_{2}\right) + \Delta_{2}^{\prime}cos\left(p_{2} \right) \\
     \Delta_{1}^{\prime}cos\left(k_{3}\right) + \Delta_{2}^{\prime}cos\left(p_{3} \right) &0  & \Delta_{1}^{\prime}cos\left(k_{3}\right) + \Delta_{2}^{\prime}cos\left(p_{3} \right) \\
     \Delta_{1}^{\prime}cos\left(k_{2}\right) + \Delta_{2}^{\prime}cos\left(p_{2} \right) &\Delta_{1}^{\prime}cos\left(k_{1}\right) + \Delta_{2}^{\prime}cos\left(p_{1} \right) &0
    \end{pmatrix}
\end{equation}  
\end{widetext}

with 
\begin{equation}
   \Delta_{1}^{(0)} = J_{1}\left[1-3sin^{2}\left(\eta \right) \right]-\sqrt{3}\left[D_{z}cos^{2}\left( \eta\right)+D_{p}sin\left(2\eta \right) \right] 
\end{equation}
\begin{equation}
    \Delta_{2}^{(0)} = J_{2}\left[1-3sin^{2}\left(\eta \right)\right]
\end{equation}
\begin{align}
    &\Delta_{1} = \frac{1}{2}\left\lbrace\left[1-3sin^{2}\left( \eta \right) \right]J_{1} + \sqrt{3}\left[ 1+ sin^{2}\left(\eta \right) \right]D_{z} \right. \notag \\
    &\left.-\sqrt{3}sin\left( 2\eta\right)D_{p}  \right\rbrace+ i \left[cos\left(\eta \right)D_{p}+sin\left(\eta \right)\left(D_{z}+\sqrt{3}J_{1} \right) \right]
\end{align}
\begin{equation}
    \Delta_{2} = \frac{1}{2}\left[1-3sin^{2} \left( \eta\right)\right]J_{2} + i\sqrt{3}sin\left(\eta \right)J_{2}
\end{equation}

 \begin{equation}
     \Delta_{1}^{\prime} = \frac{1}{2}\left[ cos^{2}\left(\eta \right)\left(\sqrt{3}D_{z} -3J_{1} \right) + \sqrt{3}sin\left(2\eta \right)D_{p} \right]
 \end{equation}

\begin{equation}
    \Delta_{2}^{\prime} = -\frac{3}{2}cos^{2}\left(\eta \right)J_{2}
\end{equation}
and using the abbreviated notations $k_{i} = \boldsymbol{k}\cdot \boldsymbol{e}_{i}$ and $p_{i} = \boldsymbol{k}\cdot\boldsymbol{e}_{i}^{\prime}$ where $\boldsymbol{e}_{1} = \left(-\frac{1}{2},-\frac{\sqrt{3}}{3} \right)$; $\boldsymbol{e}_{2} = \left(1,0 \right)$; $\boldsymbol{e}_{3} = \left(-\frac{1}{2},\frac{\sqrt{3}}{2} \right)$; $\boldsymbol{e}_{1}^{\prime} = \boldsymbol{e}_{2}- \boldsymbol{e}_{3}$; $\boldsymbol{e}_{2}^{\prime} = \boldsymbol{e}_{3}-\boldsymbol{e}_{1}$; $\boldsymbol{e}_{3}^{\prime}=\boldsymbol{e}_{1}-\boldsymbol{e}_{2}$.

\begin{figure*}
\centering
\includegraphics[width=0.85\textwidth]{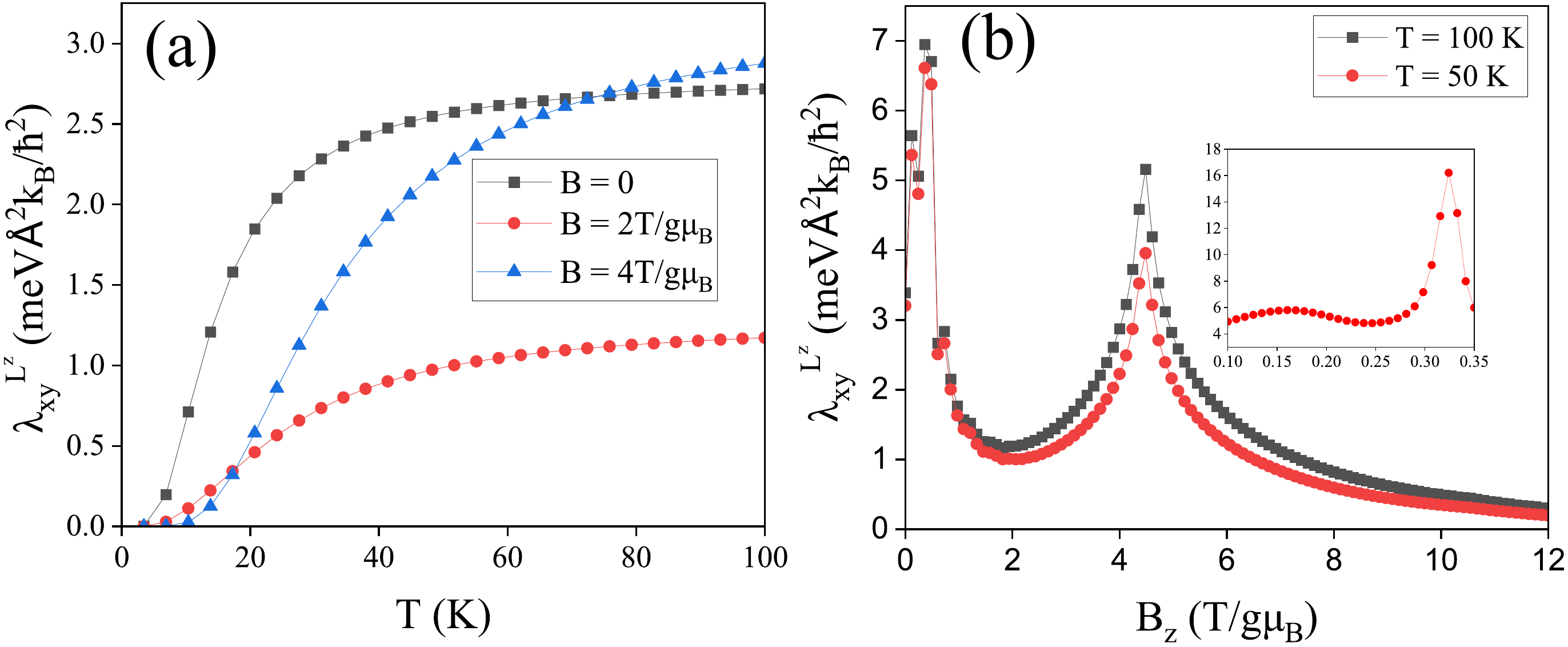}
\caption{The magnon orbital Nernst conductivity in KFe$_{3}$(OH)$_{6}$(SO$_{4}$)$_{2}$ as a function of (a) temperature T and (b) externally applied magnetic field. The calculations are performed for (a) three different applied magnetic fields and (b) two different temperatures as indicated in the legend. The inset in (b) provides a magnified view of the two peaks near $B_{z} = 0$.}
\label{FIG4}
\end{figure*}

Diagonalizing the boson Hamiltonian \eqref{Hsecond} yields the magnon dispersion $E_{n}\left( \boldsymbol{k}\right)$ \cite{Colpa1978,Colpa1986a,Colpa1986b}. Notably, neutron scattering is commonly used in experiments to investigate the magnon spectrum \cite{Bayrakci2013, Mourigal2013, Princep2017, Shamoto2018,Morano2024, McClarty2024}. The results obtained with neutron scattering are directly related to the magnon spectral function, which we can compute from the dispersion relation $E_{n}\left( \boldsymbol{k}\right)$. Specifically, the magnon spectral function is defined as \cite{Bajpai2021,Garcia2024}:
\begin{equation}\label{eq:spectralf}
 \mathcal{A}(\mathbf{k},E) = -\frac{1}{\pi} \mathrm{Im}\, G(\mathbf{k},E),
 \end{equation}
where $G(\mathbf{k},E)$ represents the one-particle retarded Green’s function calculated within the framework of quantum many-body theory, given by:
\begin{equation}\label{eq:gf}
G^{-1}\left(\mathbf{k},E\right) = E_{0} - E(\mathbf{k}) -\Sigma(\mathbf{k},E),
\end{equation}
Here $\Sigma(\mathbf{k},E)$, the self-energy, encapsulates interaction effects. The real part of $\Sigma(\mathbf{k},E)$ renormalizes the magnon energy $E(\mathbf{k}$), while its imaginary part accounts for the finite lifetime of the magnon states \cite{Bayrakci2013}. The self-energy, $\Sigma(\mathbf{k},E)$, can be computed in several ways. A perturbative approach involves evaluating specific Feynman diagrams containing loops. Alternatively, nonperturbative methods can be employed to compute the exact Green's function in Eq.~\eqref{eq:gf} for spin Hamiltonians such as the one we use here. 

In Fig.~\ref{FIG1}(a) we present the spectral function $\mathcal{A}\left(\boldsymbol{k}, E\right)$ calculated nonperturbatively for KFe$_{3}$(OH)$_{6}$(SO$_{4}$)$_{2}$ using the Hamiltonian in Eq.~\eqref{Hamilton}. We then incorporate the effect of the self-energy $\Sigma(\mathbf{k},E)$ from Eq.\eqref{eq:gf} by introducing a broadening parameter $\kappa = Im\left[\Sigma(\mathbf{k},E) \right] \approx \frac{\hbar}{2\tau}$ to account for the magnon lifetime $\tau$. The magnetic interaction parameters used in this calculations for KFe$_{3}$(OH)$_{6}$(SO$_{4}$)$_{2}$ are $S=5/2$; $J_{1}=3.18~meV$; $J_{2}=0.11~meV$; $D_{p}=0.062J_{1}$; $D_{z}=-0.062J_{1}$ \cite{Laurell2018}. The spectral function  $\mathcal{A}\left(\boldsymbol{k}, E\right)$ illustrates the magnon dispersion in KFe$_{3}$(OH)$_{6}$(SO$_{4}$)$_{2}$ along the high-symmetry path $\Gamma-K-M-\Gamma$. Notably, there are small anticrossing gaps between the lowest and middle, and between middle and  magnon bands along the $\Gamma-K$ path (see green circle). However, these gaps is not visible because the anticrossing energy gaps are comparable in magnitude to the linewidth that comes from the finite lifetime broadening. This suggests that observing this gap in neutron scattering experiments may be challenging. Although these gap are barely discernible in Fig.\ref{FIG1}(a), their presence is crucial for understanding the topological nature of the magnon bands, as will be demonstrated in the subsequent discussion.

In addition to the eigenvalues that generate the magnon dispersion, one also obtains the eigenvectors $\left\vert n \left( \boldsymbol{k} \right)\right\rangle$, which serve as the foundation for computing the dynamical properties of magnons in the system, as detailed in the following discussion.

\subsection{Berry curvature and Chern number}
Magnons are charge-neutral quasi-particles that can be generated by introducing a temperature gradient in a magnetic system. The propagation of a magnon within the system can be visualized as the motion of a wave packet, governed by both the magnon's dispersion relation and the Berry curvature of the magnon Bloch band. Specifically, considering a magnon wave packet localized around a center determined by $r_{c}$ and with a wave vector $k_{c}$, the dynamics of the magnon wave packet in the semiclassical framework are described by the following equation \cite{Matsumoto2011a,Matsumoto2011b}:
 
\begin{equation}
    \dot{\boldsymbol{r}} = \frac{1}{\hbar}\frac{\partial E_{n}\left(\boldsymbol{k} \right)}{\partial \boldsymbol{k}} - \dot{\boldsymbol{k}} \times \boldsymbol{\Omega}_{n} \left(\boldsymbol{k} \right)
    \label{velocity}
\end{equation}
with 
\begin{equation}
    \hbar \dot{\boldsymbol{k}} = -\boldsymbol{\nabla}U\left(\boldsymbol{r} \right)
\end{equation}
where $U\left(\boldsymbol{r} \right)$ is the confinement potential that is present only near the boundary of the system and $\boldsymbol{\Omega}_{n}\left( \boldsymbol{k}\right)$ the Berry curvature of the nth magnon band given by
\begin{align} \label{EqBerry}
   \boldsymbol{\Omega}_{n}\left(\boldsymbol{k} \right) = \sum_{m\neq n}\frac{i\hbar^{2}\sigma_{3}^{nn}\sigma_{3}^{mm}\langle n(\boldsymbol{k})\vert \hat{\boldsymbol{v}}_{\boldsymbol{k}} \vert m(\boldsymbol{k}) \rangle \times \langle m(\boldsymbol{k})\vert \hat{\boldsymbol{v}}_{\boldsymbol{k}} \vert n(\boldsymbol{k}) \rangle}{\left[ \sigma_{3}^{nn}E_{n} \left( \boldsymbol{k} \right) - \sigma_{3}^{mm}E_{m} \left( \boldsymbol{k} \right) \right]^{2}}.
\end{align}
Here $\left\vert n \left( \boldsymbol{k} \right) \right\rangle$ and $E_{n}\left(\boldsymbol{k} \right)$ are the eigenvectors and eigenvalues of Hamiltonian \eqref{Hsecond}, $\hat{\boldsymbol{v}}_{\boldsymbol{k}} = \frac{1}{\hbar}\frac{\partial H_{\boldsymbol{k}}}{\partial \boldsymbol{k}}$ is the velocity operator, and
\begin{equation}
    \sigma_{3} = \begin{pmatrix}
    \boldsymbol{1}_{N \times N} & 0 \\
    0 & -\boldsymbol{1}_{N \times N}
    \end{pmatrix}
\end{equation}
with $\boldsymbol{1}_{N \times N}$ a $N \times N$ identity matrix and $\sigma_{3}^{nn}=\langle n(\boldsymbol{k})\vert \sigma_{3}\vert n(\boldsymbol{k})\rangle$ the nth diagonal element of $\sigma_{3}$.

The velocity $\dot{\boldsymbol{r}}$ of a magnon wave packet comprises two distinct components. The first is the conventional group velocity, while the second corresponds to an anomalous velocity associated with the Berry curvature. The anomalous magnon current density arising from this second term is expressed as:
\begin{align}
    \boldsymbol{j}_{A} = \frac{1}{V\hbar} \sum_{n,\boldsymbol{k}} \rho\left[E_{n}\left(\boldsymbol{k} \right), T \right]\boldsymbol{\nabla}U\left(\boldsymbol{r} \right)\times \boldsymbol{\Omega}_{n} \left(\boldsymbol{k} \right)
    \label{anomalouscurrent}
\end{align}
Here
\begin{equation}
    \rho\left[E_{n}\left(\boldsymbol{k} \right), T \right] = \frac{1}{e^{E_{n}\left(\boldsymbol{k} \right)/k_{B}T}-1}
\end{equation}
is the Bose-Einstein distribution function.

Supposing that the confinement potential $U\left(\boldsymbol{r} \right)$ is $\boldsymbol{k}$-independent, one can see that under the time-reversal symmetry (TRS) operation the Berry curvature transforms as $\boldsymbol{\Omega}_{n} \left(\boldsymbol{k} \right) \rightarrow -\boldsymbol{\Omega}_{n} \left(-\boldsymbol{k} \right)$. Meanwhile, under inversion symmetry the Berry curvature transforms as $\boldsymbol{\Omega}_{n} \left(\boldsymbol{k} \right) \rightarrow \boldsymbol{\Omega}_{n} \left(-\boldsymbol{k} \right)$. Therefore, in a system possessing both TRS and inversion symmetry the Berry curvature vanishes identically throughout the entire Brillouin zone, resulting in a zero anomalous magnon current density.

To generate a nonzero anomalous magnon current, the breaking of TRS and inversion symmetry is required. Because the confinement potential $U\left(\boldsymbol{r} \right)$ in Eq.~\eqref{anomalouscurrent} is finite only near the edges of the system, the anomalous magnon current flows along these edges in the form of magnon edge states. The number of magnon edge modes is determined by the Chern number, which is defined for the nth band as follows \cite{ Mook2014a,Mook2014b, Mook2015}
\begin{equation}
    C_{n} = \frac{1}{2\pi}\int_{BZ}dk^{2}\Omega_{n}\left(\boldsymbol{k} \right)
\end{equation}

Figs.\ref{FIG1}(b-d) show the Berry curvature across the entire Brillouin zone for all three magnon modes in KFe$_{3}$(OH)$_{6}$(SO$_{4}$)$_{2}$, ordered by energy from the lowest (b) to the highest (e) band. The Berry curvature of these individual bands exhibits the $C_{3}$ symmetry characteristic of the Kagom\'e lattice illustrated in Fig.\ref{Scheme}(a). Furthermore, the topology of all three bands is nontrivial, with finite Chern number for each band. Specifically, the Chern numbers are $C = -3$, $C = 1$, and $C = 2$ for the lowest, middle, and upper bands, respectively. As expected for bosonic systems such as magnons, the sum of the Chern numbers satisfies $\sum_{n}C_{n}=0$. We note that this nontrivial topology of magnon bands in KFe$_{3}$(OH)$_{6}$(SO$_{4}$)$_{2}$ arises from the Dzyaloshinskii-Moriya interaction (DMI). Specifically, the DMI induces hybridization between magnon bands with different characteristics, leading to the opening of gaps at the $\Gamma$ and M points as well as between the lowest and middle bands along the $\Gamma-K$ direction as indicated by the green circle in Fig.\ref{FIG1}(a). In the absence of DMI all the magnon bands are topologically trivial. In the next part we will see that the closing and reopening of these gap leads to a topological transition of the magnon system in KFe$_{3}$(OH)$_{6}$(SO$_{4}$)$_{2}$ under an applied magnetic field.

We emphasize that magnons exhibit behavior distinct from that in fermionic systems. For fermionic systems such as electrons in a topological insulator the nontrivial topology manifests as an insulating bulk and conducting boundary states. For bosonic systems such as magnons, at finite temperatures all bands contribute to magnon transport. Consequently, while edge states arising from the nontrivial topological bands of magnons do exist, they cannot be clearly distinguished from bulk states through, for example, the thermal Hall conductivity of magnons. In other words, the magnon group velocity, as described in Eq.~\eqref{velocity}, remains finite at finite temperatures, leading to magnon current density distributed throughout the material rather than being confined to the edges. Consequently, and unlike electrons in topological insulators, magnons do not exhibit quantized conduction.

\subsection{Magnon Orbital moment and magnon Orbital Nernst effect}
Applying a temperature gradient along the $\hat{x}$ direction to a magnetic material like KFe$_3$(OH)$_6$(SO$_4$)$_2$ can induce magnon propagation within the system. Because magnons carry spin angular momentum, the presence of DMI gives rise to a transverse ($\hat{y}$) spin current carried by magnons. This phenomenon is known as the magnon spin Nernst effect. In addition to spin, magnons also possess orbital angular moment (OAM) due to the intrinsic self-rotation of magnon wave packets. Consequently, the temperature gradient can also induce a magnon orbital Nernst effect, even in the absence of DMI. Within the linear response theory, one can derive the linear thermal response mechanism governing the flow of magnon OAM to be \cite{To2025} 
\begin{align}
  j^{L^{z}}_{y} = -\lambda ^{L^{z}}_{xy}\partial_{x}T  
\end{align}
where
\begin{align}
  \lambda_{xy}^{L^{z}} = \frac{2k_{B}}{\hbar V} \sum_{\boldsymbol{k}}\sum_{n=1}^{N} \Omega_{xy}^{L^{z},n}\left(\boldsymbol{k} \right)F\left( \rho\left[E_{n}\left(\boldsymbol{k} \right), T \right]\right)  
\end{align}
is the total Orbital Nernst conductivity. Here
\begin{equation}
    F\left( \rho\right) = \left(1+ \rho \right)ln\left(1+ \rho \right) -\rho ln\left(\rho \right)
    \label{F1}
\end{equation}
and the total Orbital Berry curvature is given by 
\begin{widetext}
     \begin{align}
   \Omega^{L^{z},n}_{xy}\left(\boldsymbol{k} \right) =-\sum_{m \neq n}2 \hbar^{2}\sigma_{3}^{nn}\sigma_{3}^{mm}Im \left\lbrace\frac{\left\langle n\left(\boldsymbol{k} \right) \left\vert    \hat{j}^{L^{z}}_{x}    \right\vert m\left(\boldsymbol{k} \right) \right\rangle   \left\langle m\left(\boldsymbol{k} \right) \left\vert \hat{v}_{y}\right\vert n\left(\boldsymbol{k} \right) \right\rangle}{\left(\left[\boldsymbol{\sigma}_{3}E\left(\boldsymbol{k} \right) \right]_{mm} - \left[\boldsymbol{\sigma}_{3}E\left(\boldsymbol{k} \right) \right]_{nn}\right)^{2}}\right\rbrace 
\end{align}   
\end{widetext}
where $\boldsymbol{j}^{\hat{L}^{z}} = \hat{L}^{z}\boldsymbol{\sigma}_{3}\hat{\boldsymbol{v}}+\hat{\boldsymbol{v}}\boldsymbol{\sigma_{3}}\hat{L}^{z}$ is the orbital current operator with $\hat{L}^{z}$ the z-component of the orbital moment operator of the magnon:
\begin{align}
    \hat{\boldsymbol{L}}=\frac{1}{4}\left(\hat{\boldsymbol{r}} \times \hat{\boldsymbol{v}} - \hat{\boldsymbol{v}} \times \hat{\boldsymbol{r}} \right)
    \label{Loper}
\end{align}
Here $\hat{\boldsymbol{r}}$ and $\hat{\boldsymbol{v}}$ are the position operator and the velocity operator in real space, respectively. The matrix elements of the orbital angular moment operator are given by \cite{To2025}:
\begin{align}
    \boldsymbol{L}_{mn}\left(\boldsymbol{k}\right)=\left\langle m\left(\boldsymbol{k} \right)\left\vert \hat{\boldsymbol{L}}  \right\vert  n\left(\boldsymbol{k}\right)\right\rangle =-i\hbar \boldsymbol{\mathcal{N}}_{mn}
\end{align}
where 
\begin{widetext}
  \begin{align}
    \boldsymbol{\mathcal{N}}_{mn}=\frac{1}{4} \sum_{p\neq m,n}\sigma_{3}^{pp}&\left(\frac{1}{\left[ \boldsymbol{\sigma}_{3}E \left(\boldsymbol{k} \right) \right]_{mm} - \left[ \boldsymbol{\sigma}_{3}E\left(\boldsymbol{k} \right) \right]_{pp}} + \frac{1}{\left[ \boldsymbol{\sigma}_{3}E\left(\boldsymbol{k} \right) \right]_{nn} - \left[ \boldsymbol{\sigma}_{3}E\left(\boldsymbol{k} \right) \right]_{pp}}\right) \left\langle m\left(\boldsymbol{k} \right)\left\vert \hat{\boldsymbol{v}}_{\boldsymbol{k}}\right\vert p \left( \boldsymbol{k} \right)\right\rangle \times \left\langle p\left(\boldsymbol{k} \right)\left\vert \hat{\boldsymbol{v}}_{\boldsymbol{k}}\right\vert n \left( \boldsymbol{k} \right)\right\rangle. 
    \label{Ntensor}
\end{align}  
\end{widetext}

$L^{z}_{nn}$ is the z-component of the intra-band magnon OAM in the Bloch wave of the n$^{th}$ band given by
\begin{align} \label{OBM}
    \boldsymbol{L}_{nn}\left(\boldsymbol{k} \right)= -\frac{i\hbar}{2}\sum_{p\neq n}\sigma_{3}^{pp}\frac{ \left\langle n\left(\boldsymbol{k} \right)\left\vert \hat{\boldsymbol{v}}_{\boldsymbol{k}}\right\vert p \left( \boldsymbol{k} \right)\right\rangle \times \left\langle p\left(\boldsymbol{k} \right)\left\vert \hat{\boldsymbol{v}}_{\boldsymbol{k}}\right\vert n \left( \boldsymbol{k} \right)\right\rangle }{ \sigma_{3}^{nn}E_{n}\left(\boldsymbol{k} \right)  - \sigma_{3}^{pp}E_{p}\left(\boldsymbol{k} \right) }.  
\end{align}

In Fig.~\ref{FIG2}, we present the intra-band magnon orbital angular moment (a-c) and the orbital Berry curvature of the magnon bands (d-f) in KFe$_{3}$(OH)$_{6}$(SO$_{4}$)$_{2}$, ordered from the lowest (a,d) to the highest (c,f) energy of the bands as depicted in Fig.~\ref{FIG1}(a). Importantly, the intra-band magnon OAM and orbital Berry curvature of the lowest band are an  order of magnitude smaller than those of the middle and highest bands, as indicated by the color coding axes in Fig.~\ref{FIG2}. This suggests that the contribution of the lowest band to the orbital Nernst effect of magnons in KFe$_{3}$(OH)$_{6}$(SO$_{4}$)$_{2}$ is relatively minor compared to that of the higher bands. This, in turn, suggests that the topological phase transition of the two upper bands (middle and highest) may have a greater influence on the magnon orbital Nernst effect than the topological transition of the two lower bands (middle and lowest). This will be examined in detail in the next section, which focuses on the topological phase transitions of the magnonic system in KFe$_{3}$(OH)$_{6}$(SO$_{4}$)$_{2}$ under an externally applied out-of-plane magnetic field.

\begin{figure}
\centering
\includegraphics[width=0.5\textwidth]{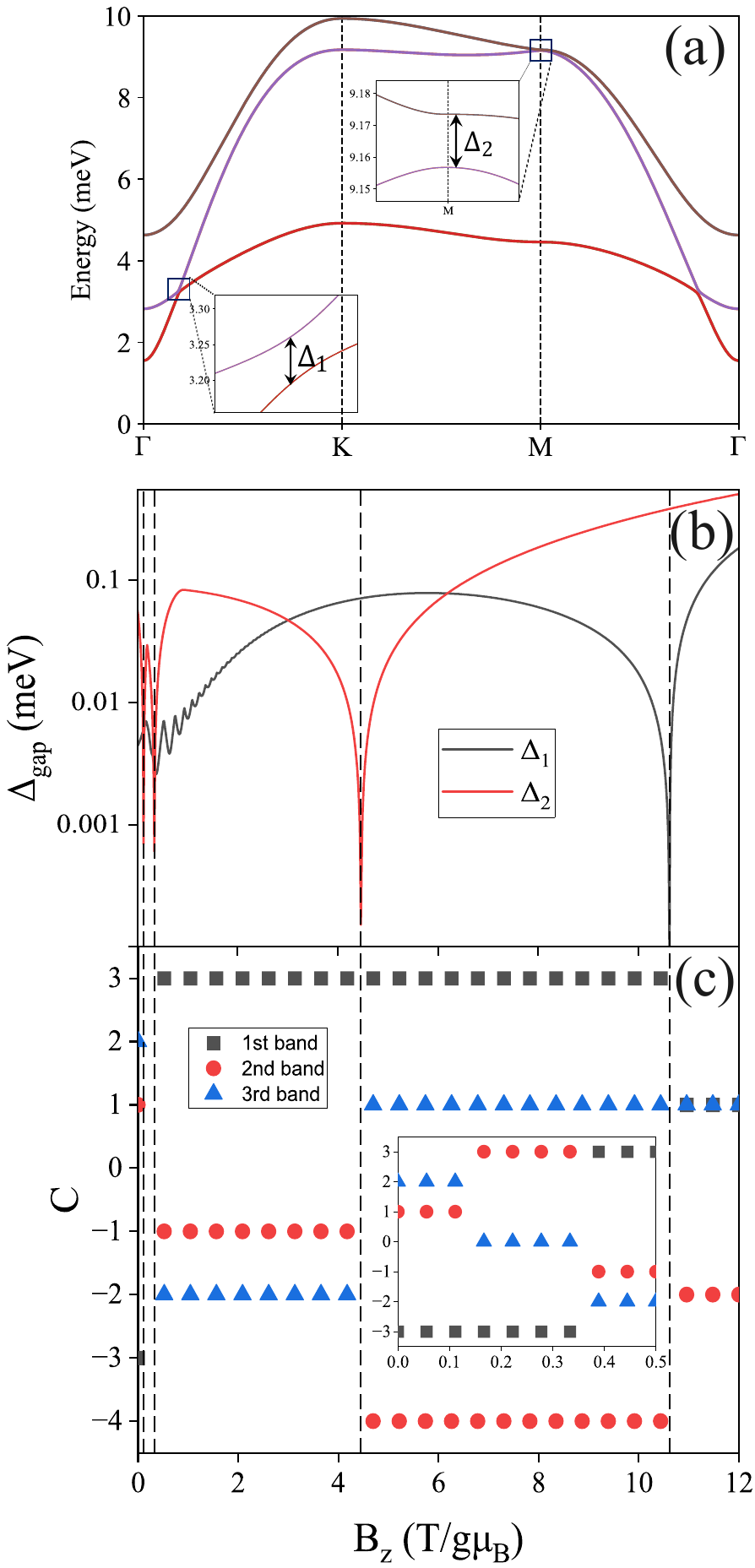}
\caption{(a) Magnon dispersion in the noncollinear AFM KFe$_{3}$(OH)$_{6}$(SO$_{4}$)$_{2}$ along the $\Gamma-K-M-\Gamma$ path calculated for an applied magnetic field $B = 4~T/g\mu_{B}$. The gap between the middle and lowest magnon bands ($\Delta_{1}$) and between the highest and the middle magnon bands ($\Delta_{2}$) are indicated by rectangles with insets showing the gaps quantitatively. The variation of (b) the gaps $\Delta_{1}$ and $\Delta_{2}$ and (c) the Chern numbers of the three magnon bands (see panel (a)) as a function of the externally applied magnetic field. The inset in (c) presents a zoomed-in view of the Chern number at low magnetic fields, ranging from 0 to 0.5~$T/g\mu_{B}$.}
\label{FIG3}
\end{figure}

\section{Magnon Orbital Nernst effect in Potassium iron jarosite} \label{MONE}
We now turn to an investigation of the magnon ONE along the $\hat{y}$-direction in potassium iron jarosite [KFe$_{3}$(OH)$_{6}$(SO$_{4}$)$_{2}$] under a temperature gradient in the $\hat{x}$-direction and an applied out-of-plane magnetic field along the $\hat{z}$-direction ($B_{z}$). In Fig.~\ref{FIG4}(a), we present the Orbital Nernst conductivity (ONC) of magnons in  KFe$_{3}$(OH)$_{6}$(SO$_{4}$)$_{2}$ as a function of the system's average temperature, T, under varying magnitudes of an externally applied magnetic field $B_{z}$. As shown by the black points in Fig.~\ref{FIG4}(a), the ONC increases with temperature, which can be attributed to the enhanced contribution of higher magnon bands to the transport properties of the system as the temperature increases. In the absence of an external magnetic field, the ONC saturates when the temperature reaches approximately 40 K ($k_{B}T \approx 3.4~meV$), coinciding with the energy at which the highest magnon band [$\sim$3.5 meV at the $\Gamma$-point, see Fig.~\ref{FIG1}(a)] comes into play. Applying an external magnetic field of $2T/g\mu_{B}$ shifts the magnon band energies upward and results in both a decrease in the ONC across the entire temperature range and an increase in the temperature at which the ONC saturates, as shown by the red points in Fig.\ref{FIG4}(a). Increasing the magnetic field from $2T/g\mu_{B}$ to $4T/g\mu_{B}$ further increases the ONC saturation temperature, as shown by the blue points in Fig.\ref{FIG4}(a), but it also \textit{increases} the ONC relative to the case at $2T/g\mu_{B}$ for temperatures above $\sim$20~K. Indeed, the ONC under a $4T/g\mu_{B}$ magnetic field also exceeds the ONC of the $B_{z} = 0$ case at temperatures above 70~K. This behavior is more clearly illustrated in Fig.~\ref{FIG4}(b), which presents the ONC as a function of the magnetic field for two average temperatures, T. Initially, increasing the applied magnetic field from $B_{z} = 0$ increases the ONC, with prominent maxima at $B_{z}\approx 0.15 T/g\mu_{B}$ and $0.4 T/g\mu_{B}$ [see the inset in Fig.\ref{FIG4}(b)]. For all of the temperatures we studied, we find that the ONC at $B_{z} = 0.4 T/g\mu_{B}$ is approximately five times greater than the ONC in the absence of a magnetic field. As the magnetic field increases beyond $B_{z} = 0.4 T/g\mu_{B}$ the ONC decreases, reaching a local minimum at $B \approx 2 T/g\mu_{B}$ before rising again to a local maximum at $B_{z} \approx 4.4 T/g\mu_{B}$. After the maximum at $\approx 4.4 T/g\mu_{B}$, the ONC gradually declines. Notably, the behavior of the ONC as a function of magnetic field is similar for both $T=50K$ and $T = 100K$, underscoring the robustness of this phenomenon, whose origins we will now explain.

The phase transitions in the ONC at $B_{z}\approx 0.15$, $0.4$, and $4.4~T/g\mu_{B}$ in Fig.\ref{FIG4}(b) correspond to topological phase transitions in the magnon bands induced by variations in the applied magnetic field. The topological phase transitions are quantitatively captured by the changes in the Chern number of each band as a function of the magnetic field, as shown in Fig.\ref{FIG3}(c). These transitions result from changes in anticrossings of distinct magnon bands due to the applied magnetic field. The three magnon bands are shown in Fig.\ref{FIG3}(a), the anticrossing points are highlighted by rectangles, and the magnitude of the anticrossing gaps are shown in the insets. The externally applied magnetic field not only shifts the magnon energy levels in the system, it also acts as an effective Dzyaloshinskii-Moriya interaction. It is the changing contribution of this effective DMI that causes the changes in the anticrossings between magnon bands and the resulting topological phase transitions. Specifically, Eq.~\ref{cantingangle} shows that varying the magnetic field modifies the canting angle, which, in turn, alters both the in-plane and out-of-plane components of the DMI. This variation alters the strength of the magnon band hybridization that determines the magnitude of the anticrossing energy gaps. Consequently, the gaps between the middle and lowest magnon bands ($\Delta_{1}$) and between the highest and middle magnon bands ($\Delta_{2}$) close and reopen as the magnon energy levels and hybridization strength vary with the magnetic field. To show this behavior more clearly, Fig.\ref{FIG3}(b) plots the gap between the middle and lowest magnon branches ($\Delta_{1}$) and the gap between the upper and middle branches ($\Delta_{2}$) as functions of $B_{z}$. The gap $\Delta_{2}$ closes and reopens three times, at $B_{z}\approx 0.15$, $0.4$, and $4.4~T/g\mu_{B}$. A similar collapse of the anticrossing energy gap occurs for $\Delta_{1}$ when $B_{z}$ crosses $10.5~T/g\mu_{B}$. These transitions explain the topological phase changes that are shown by the discrete change in the Chern number of each band [Fig.~\ref{FIG3}(c)]. Specifically, the Chern numbers of the three bands change from (-3,1,2) to (-3,3,0) at $B_{z}\approx 0.15 T/g\mu_{B}$, from (-3,3,0) to (3,-1,-2) at $B_{z} \approx 0.4 T/g\mu_{B}$, from (3,-1,-2) to (3,-4,1) at $B_{z} \approx 4.4 T/g\mu_{B}$, and from (3,-4,1) to (1,-2,1) at $B_{z} \approx 10.5 T/g\mu_{B}$. The topological phase transition caused by the closing and reopening of the gap between the highest and middle magnon bands ($\Delta_{2}$) is the most important for the affects we discuss here because these bands dominate the contribution to the orbital Nernst effect, as discussed earlier.

Finally, we note that the topological phase transitions, marked by the closing and reopening of gaps between magnon bands, result from the interplay between a) the DMI and b) the Zeeman interaction between the local magnetic moments and the externally applied magnetic field. In the absence of DMI, all magnon bands become topologically trivial and the influence of the externally applied magnetic field on magnon orbital moment transport is weak (data not shown). This behavior is akin to that observed in collinear antiferromagnets on a honeycomb lattice even when DMI is present. Consequently, in noncollinear antiferromagnets such as KFe$_{3}$(OH)$_{6}$(SO$_{4}$)$_{2}$ DMI plays a pivotal role in enabling the tunability of the magnon Orbital Nernst effect and the topological properties of magnon bands under the influence of an externally applied magnetic field. These findings underscore the essential role of DMI in facilitating the control of magnon-based transport phenomena, offering promising prospects for manipulating topological magnonic orbital moment properties in noncollinear antiferromagnetic materials.

\section{Conclusion and discussion}\label{Conclusion}
Our theoretical investigation reveals the tunable intrinsic magnon orbital Nernst effect in noncollinear antiferromagnets with Kagom\'e spin systems. We demonstrate that an externally applied magnetic field induces topological phase transitions in the magnonic band structure, characterized by the closing and reopening of the anticrossing energy gap at critical points in the Brillouin zone. These transitions underscore the tunability of orbital Nernst conductivity via magnetic field modulation.

These results are of particular relevance to experimental probes of the magnon ONE because the presence of orbital Nernst currents leads to the accumulation of magnon orbital moments at the system’s edges. Under the influence of the Dzyaloshinskii-Moriya interaction, this accumulation manifests as orbital magnetization of magnons, a measurable quantity detectable via the magneto-optical Kerr effect. This phenomenon, first theoretically predicted by R. R. Neuman et al. \cite{Neumann2020}, has been experimentally observed in several systems \cite{Zhang2020im, Alahmed2022, Huang2024}.  Moreover, the interplay between magnon spin currents and magnon orbital moment-induced electric polarization can generate a transverse voltage \cite{To2025}. The findings reported here indicate that the magnon ONE can be significantly enhanced by applying an external magnetic field. These insights thus provide a framework for enhancing the magnitude of experimentally measurable magnonic effects, paving the way toward advanced control of magnonic phenomena in noncollinear antiferromagnets.

\begin{acknowledgments}
This research was primarily supported by NSF through the University of Delaware Materials Research Science and Engineering Center, DMR-2011824.
\end{acknowledgments}

\bibliography{Ref}





\end{document}